# Anisotropic band flattening in graphene with 1D superlattices

Yutao Li[1], Scott Dietrich[1,2], Carlos Forsythe[1,3], Takashi Taniguchi[4], Kenji Watanabe[4], Pilkyung Moon[5,6], Cory R. Dean[1]

[1] Department of Physics, Columbia University, New York, NY, USA
[2] Current address: Department of Physics, Villanova University, Villanova, PA, USA.
[3] Current address: Molecular Foundry, Lawrence Berkeley National Lab, Berkeley, CA, USA
[4] National Institute for Materials Science, Tsukuba, Japan
[5] New York University Shanghai and NYU-ECNU Institute of Physics at NYU Shanghai, Shanghai, China
[6] State Key Laboratory of Precision Spectroscopy, East China Normal University, Shanghai, China

**Patterning graphene with a spatially-periodic potential provides a powerful means to modify its electronic properties. [1-3] Dramatic effects have been demonstrated in twisted bilayers where coupling to the resulting moiré-superlattice yields an isolated flat band that hosts correlated many-body phases.[4, 5] However, both the symmetry and strength of the effective moiré potential are constrained by the constituent crystals, limiting its tunability. Here we exploit the technique of dielectric patterning[6] to subject graphene to a one-dimensional electrostatic superlattice (SL)[1]. We observe the emergence of multiple Dirac cones and find evidence that with increasing SL potential the main and satellite Dirac cones are sequentially flattened in the direction parallel to the SL basis vector. Our results demonstrate the ability to induce tunable transport anisotropy in high mobility two-dimensional materials, a long-desired property for novel electronic and optical applications[7], as well as a new approach to engineering flat energy bands where electron-electron interactions can lead to emergent properties[8].**

Two-dimensional (2D) materials such as graphene and the transition metal dichalcogenides (TMDCs) exhibit a wide range of electronic and optical properties, making them ideal platforms for both exploring fundamental phenomenon and promising building blocks for the next generation devices[9, 10]. Imposing spatially-periodic external fields has proven to be a powerful technique to gain further control over the opto-electronic response in these materials, since the interplay between the externally applied superlattice potential, and the intrinsic lattice site potentials of the 2D crystal, can be exploited to modify the electronic bandstructure[4, 6, 11-14]. In the case of graphene, this has enabled realization of a variety of new phenomenon that are not inherent in the native material including additional Dirac cones[6, 11, 12, 15, 16], superconductivity[4, 13], Mott-like insulating states[5], and the appearance of ferromagnetic ordering and topologically non-trivial subbands[14, 17].

So far, most experimental efforts to pattern 2D materials have focused on 2D superlattices, where the potential varies spatially along two directions. However, patterning with a one-dimensional superlattice, where the potential varies periodically along one axis only, is anticipated to give rise to a new and an equally rich variety set of electronic properties.[1, 2, 11, 18-20] In particular, 1D SL patterning of graphene is expected to yield highly anisotropic energy-momentum relation between the directions perpendicular and parallel to the SL basis vector.[1] At certain, well-defined, strengths of the SL modulation, the group velocity of charge carriers at the Dirac cone can become renormalized to zero in the direction perpendicular to the SL basis vector, leading to a "flattened" Dirac cone in one direction, while the Fermi velocity in the other direction remains unaffected[1], resulting in transport anisotropy.

Experimental studies that have investigated the transport response in graphene subject to a 1D SL potential have shown the emergence of new resistive features appearing at finite carrier density, $n$, when current was applied parallel to 1D SL basis vector.[21-25]. However, the physical origin of these resistance oscillations are unresolved and have been attributed to both Fabry-Perot interference [22-24], and band structure modifications [21, 25]. Moreover, direct observation of transport anisotropy, a key signature expected for scattering from the 1D SL, has not been experimentally demonstrated.

In this study, we exploit the technique of dielectric patterning[6] to fabricate high mobility graphene devices with a gate-tunable 1D SL. We identify SL-tunable resistance peaks as arising from satellite DPs in the bandstructure. By measuring the transport response with current applied both parallel and perpendicular to the SL, we identify a relative hierarchy of band flattening between the CNP and satellite DPs, in excellent agreement with bandstructure modelling. Our results demonstrate that 1D superlattices can be used to dynamically induce extreme anisotropy in the graphene bandstructure.

A typical device construction in our study is shown in Fig. 1a-c. Parallel straight lines with periodicity on the order of 50nm are first etched onto a thermally grown $SiO_2$/doped Si substrate. An hBN-encapsulated graphene heterostructure is then placed on top of the etched substrate using the van der Waals dry transfer technique[26] (Fig. 1a). Applying a bias to the unstructured, bottom, Si-gate translates to a spatially varying potential in the graphene layer due to the spatially varying dielectric that separates the two[6]. A separate top gate, made of unstructured few-layer graphite, is used to independently tune the average carrier density in the graphene layer (Supplementary information, S1). The device is etched into an L-shaped Hall bar, using standard nanolithography processes, and electrically contacted using the edge contact geometry[27] (Fig. 1b). The L shape allows us to measure transport response both parallel and perpendicular to the 1DSL, within the same device (Fig. 1b,c). The resistance measured with current flowing along the direction of the 1D SL basis vector (or equivalently, perpendicular to the 1D SL straight lines) is denoted as $R_{xx} = V_{xx}/I$, and the resistance measured perpendicular to the direction of the 1D SL basis vector (or equivalently, along the 1D SL straight lines) is denoted as $R_{yy} = V_{yy}/I$. (Fig. 1c).

Fig. 1d~f shows $R_{xx}$ and $R_{yy}$ versus carrier density from a device with SL period $L = 47\ nm$, measured for three different strengths of the SL modulation. $R_{xx}$ and $R_{yy}$ were measured simultaneously by sourcing a fixed current of 90 nA, through the entire device and measuring both $V_{xx}$ and $V_{yy}$. In each plot, the density was varied by tuning the top gate, $V_{TG}$, while maintaining a fixed bias on the bottom superlattice gate, $V_{SL}$. For all three values of the superlattice bias, $V_{SL}$, the resistance measured in the $yy$ direction resembles the typical response observed in unpatterned graphene, namely a single resistance peak centered at the charge neutrality point (CNP). At temperature $T = 2K$, the width of the CNP peak measures $\sim 2 \times 10^{11} cm^{-2}$, which is similar to typical values of hBN-encapsulated graphene devices with no superlattice patterning.[28] Additionally, with the top gate biased to give a carrier density of $n = 10^{12} cm^{-2}$, and the superlattice potential tuned to zero, we measured a carrier mobility of $\mu = 130,000\ cm^2/(V \cdot s)$ (See SI). This corresponds to a mean free path of ~1.5μm - similar to the device width of 2μm and an order of magnitude longer than the SL periodicity. Both metrics indicate that the superlattice patterning has not substantially degraded the graphene mobility.

In the $xx$ direction, the responses are dramatically different for each $V_{SL}$ value. At $V_{SL} = 21V$, in addition to a peak in the $R_{xx}$ at the CNP, two additional resistance peaks appear that are located symmetrically about the CNP peak (Fig. 1d). Upon increasing the superlattice gate bias to $V_{SL} = 44V$, the satellite resistance peaks grow, becoming more prominent than the CNP peak, and simultaneously shift to larger carrier density (Fig. 1e). At $V_{SL} = 88V$, the CNP becomes a resistance minimum and now there are four distinct peaks located symmetrically about the CNP (Fig. 1f). The strong asymmetry between the $xx$ and $yy$ response confirms that the 1D superlattice preferentially modifies the electron transport in the direction of the 1D superlattice wave vector .

The evolution of the $R_{xx}$ features with increasing $V_{SL}$ – i.e, the satellite resistance peaks varying in number, density position and relative magnitude, is distinctly different from the case of graphene subject to a 2D superlattice modulation[6], where by contrast the satellite resistance peak positions are fixed at all SL bias, and simply related to the SL unit cell area.

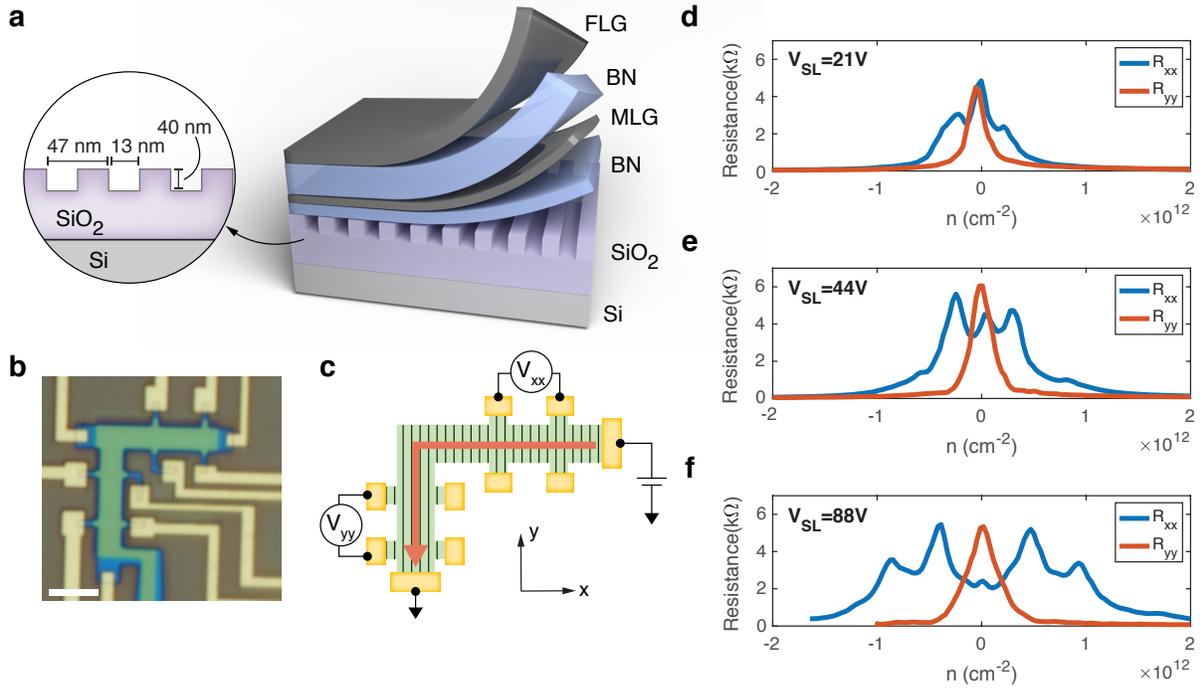

**Fig.1| Transport anisotropy in graphene subjected to a one-dimensional superlattice a**, Cartoon schematic showing our Architecture of a 1D PDSL graphene device structure. The superlattice potential arises due to patterning of the dielectric layer that separates the doped silicon substrate from the BN-encapsulated graphene device (see text) . MLG = monolayer graphene. FLG = Few Layer graphene **b**, Optical image of a 1D superlattice graphene device with period L = 47nm. Scale bar is 5μm. **c**, Schematic diagram of the L-shaped Hall bar, identifying the definitions of $V_{xx}$ and $V_{yy}$. Not to scale. Green region: graphene channel. Yellow region: Metal-to-graphene edge contacts. Parallel black lines represent the 1D SL lines etched onto SiO₂, with the line spacing greatly exaggerated. **d~f**, Resistance measurements from the $L = 47nm$ 1D SL device at three different strengths of SL modulation.

To further understand this response, we modelled the device as a graphene channel perturbed by a square-wave potential $U(x)$. The Hamiltonian of this system is given by $H = v_F \vec{\sigma} \cdot \vec{p} + I_2 V_{1D}(x)$ , where $V_{1D}(x)$ is a Kronig-Penney type potential with amplitude $V_0$ and $I_2$ is the 2x2 identity matrix. Using this model, we calculate the band structure under varying strengths of SL modulation, as quantified by the dimensionless quantity $u = \frac{V_0 L}{\hbar v_F}$ (Supplementary Information S3) where $L$ is the SL period and $v_F$ is the Fermi velocity of unpatterned graphene. Fig. 2 shows the bandstructure for $u = 2\pi, 4\pi$ and $6\pi$. The energy contour at $u = 2\pi$ shows mini Dirac cones at $E \neq 0$ at $k_x \neq 0$ (Fig. 2b, left) in addition to the main Dirac cone. At $u = 4\pi$ the main Dirac cone at $k_x = 0$ flattens in the $k_y$ direction while preserving the graphene linear dispersion relationship in the $k_x$ direction. At the same time the 1st mini-Dirac cones has linear dispersion in both $k_x$ and $k_y$ (Fig 2a,b mid). At $u = 6\pi$ the main Dirac cone restores its linear dispersion in both $k_x$ and $k_y$ directions, but now the 1st mini Dirac cone "flattens" in the $k_y$ direction, while retaining a linear dispersion in the $k_x$ direction (Fig. 2b, right). In general, we find that the $l^{th}$ mini Dirac cone ($l = 0$ corresponding to the main Dirac cone) flattens in the $k_y$ direction (but not in the $k_x$ direction) when the SL potential $u$ satisfies

$$u = \begin{cases} 4\pi N, & \text{(even } l, \text{ including } l = 0) \\ 4\pi N + 2\pi, & \text{(odd } l) \end{cases}, \qquad (1)$$

for every positive integer $N$, but excluding $u = 2\pi l$. (Fig 2c, see also Supplementary Information S3).

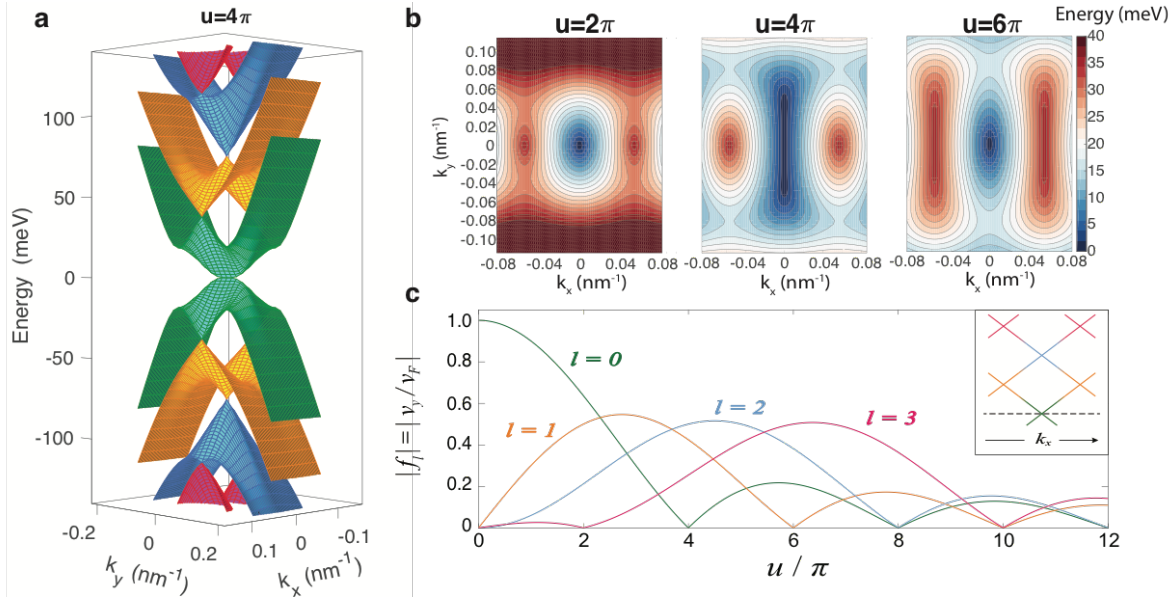

**Fig. 2|Band structure calculations for a $L = 55nm$ graphene 1D SL system.** The strength of superlattice modulation is quantified by $u = \frac{V_0 L}{\hbar v_F}$, a dimensionless quantity. **a**, Calculated band structure of a 1D SL graphene system with $u = 4\pi$. **b**, Fermi surface contours for energies between the main DP and 1$^{st}$ mini DP, for $u = 2\pi, 4\pi, 6\pi$ respectively. **c**, Normalized Fermi velocity in y direction at the $l$th Dirac cone, $|f_l| = |v_y/v_F|$, as a function of SL modulation $u$. Inset shows the band structure slice at $k_y = 0$. The main DP ($l = 0$) becomes flat in $k_y$ direction at $u = 4\pi$ and the 1$^{st}$ mini DP ($l = 0$) becomes flat in $k_y$ direction at $u = 6\pi$.

Fig. 3a shows the experimentally measured $R_{xx}$ plotted versus SL bias, $V_{SL}$, and average carrier density $n$ for a device with SL period $L = 55nm$, corresponding to the same paramaters modelled in Fig. 2. The $R_{xx}$ response exhibits an elaborate scale-like pattern of resistive maxima and minima. The number of resistive peaks increases with increasing SL bias, evolving from a single maximum at the CNP $V_{SL} = 0V$ up to 5 resolved maxima at $V_{SL} = 90V$, while at the same time each maxima shifts to higher density with increasing $V_{SL}$. Fig. 3c shows $R_{yy}$ measured from the same device. In this direction, a single peak at the CNP is observed for all values of $V_{SL}$. However, we observe that the peak magnitude first increases with increasing $V_{SL}$ and then decreases again. Fig. 3b,d show theoretical plots of $R_{xx}$ and $R_{yy}$ versus the dimensionless superlattice potential, $u$, and density, $n$, calculated from the band structure shown in Fig. 2 using relaxation time approximation[29-31]. We find excellent agreement between the measured and calculated resistance in both the *xx* and *yy* directions.

Dashed lines in Fig. 3a,b trace along features associated with the main/satellite Dirac cones. Along the main CNP peak trajectory, as $V_{SL}$ increases from 0V (corresponding to $u = 0$) to 90V (corresponding to $u \approx 7\pi$), $R_{xx}$ goes from a maximum ($u = 0$) to a minimum ($u = 4\pi$, labelled by "A") to another maximum ($u = 6\pi$). The $R_{xx}$ maxima at $u = 0$ and $u = 6\pi$ result from the low density of states at the Dirac point(s) at $E = 0$, similar to what is observed in unpatterned graphene. At $u = 4\pi$ the system is characterized by a "flattening" of the Dirac cone in the $k_y$ direction. This corresponds to a reduction of the Fermi velocity towards zero in the y-direction, i.e. a suppression of conductivity along this direction, and thus $R_{yy}$ shows a maximum. However, at the same time this band distortion results in an increase in the integrated DOS over the entire fermi surface, leading to an increased conductance, i.e. lower $R_{xx}$, measured in the x direction. Similar cycles of $R_{xx}$ maxima/minima are also observed in the satellite mini Dirac cones. For the 1st mini DP, $R_{xx}$ appears as a maximum near $u = 4\pi$, then reaches a minimum at $u = 6\pi$ (labelled by "B"), and approaches another maximum at $u = 8\pi$. The $u = 4\pi$ $R_{xx}$ maximum corresponds to the unflattened 1st mini Dirac cone (Fig. 3g) while the $u = 6\pi$ $R_{xx}$ minimum corresponds to the flattened 1st mini Dirac cone labelled by "B" in Fig. 3i. The scale-like pattern in $R_{xx}$ thus is a result of alternating flattening/unflattening of Dirac cones as the strength of SL modulation increases.

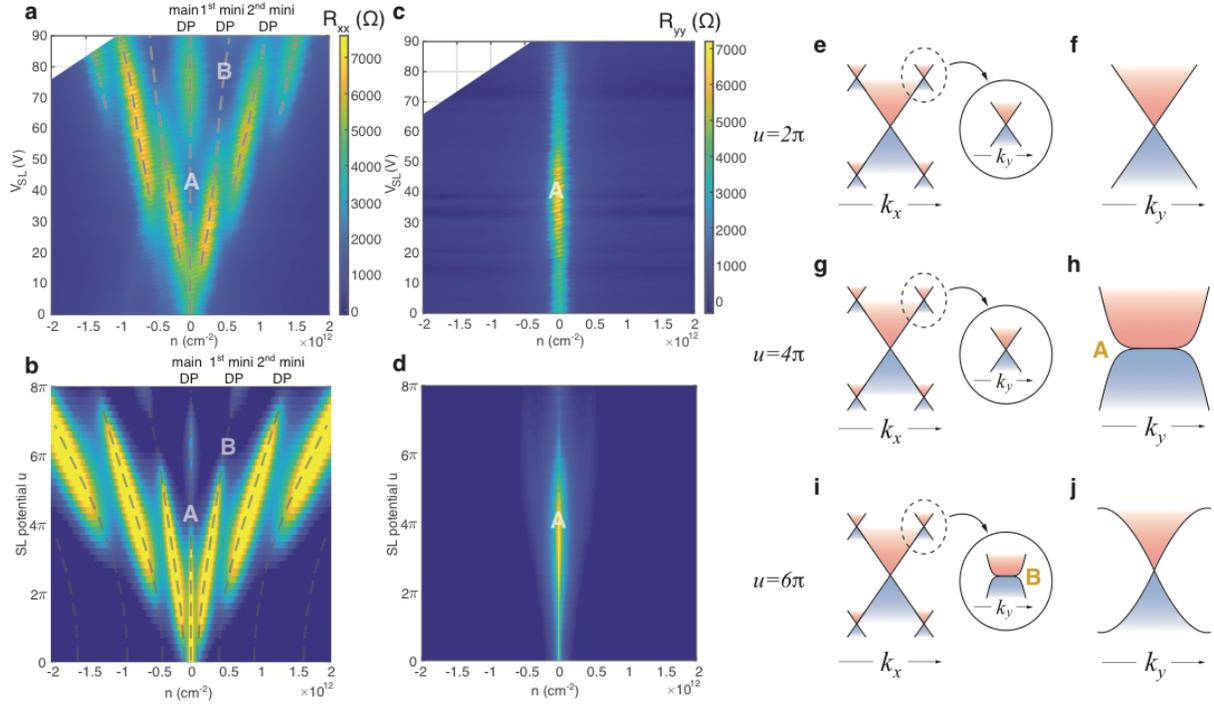

**Fig. 3| Anisotropic band flattening. a**, Device resistance $R_{xx}$ as a function of carrier density $n$ and back gate voltage $V_{SL}$, measured from an $L = 55nm$ device **b**, Calculated $R_{xx}$ for $L = 55nm$ device plotted as a function of dimensionless SL strength $u$ and carrier density $n$. In **a,b** gray dashed lines trace along features related to the same main/mini Dirac point. **c**, $R_{yy}$ measured from the same $L = 55nm$ device. **d**, Calculated $R_{yy}$ for $L = 55nm$ device **e~i**, Band structures at $u = 2\pi, 4\pi, 6\pi$ sliced along $k_y = 0$ (**e,g,i**), $k_x = 0$ (**f,h,j**), and $k_x = \pm\pi/L$ (insets**)**. Throughout this figure, select $R_{xx}$ and $R_{yy}$ extrema are related to the anisotropically flattened Dirac cones by letters A and B.

Finally, we examine magnetotransport properties in our graphene 1DSL devices. Fig. 4a,c show the longitudinal resistances $R_{xx}$ and $R_{yy}$ respectively, as a function of carrier density, $n$, and applied magnetic field $B$, for a $L = 47nm$ device. In these measurements $V_{SL} = 48V$, corresponding to $u \approx 3\pi$. $R_{xx}$ (Fig. 4a) versus magnetic field shows a Landau fan of IQHE states emanating from the CNP at $n = 0$ with filling factors identical to those of pristine graphene ($\nu = 4n + 2$, $n$ integer). In addition, two satellite fan-like features are also visible, emanating from $n = \pm 2.7 \times 10^{11}\ cm^{-2}$ (Fig. 4a, red arrows), the carrier densities at which mini Dirac cones emerge as predicted by band structure simulation in Fig. 3b. $R_{xx}$ minima features emanating from the $n = -2.7 \times 10^{11}\ cm^{-2}$ satellite fan are indicated in the fan tracing diagram shown in Fig. 4c. (Supplementary Information S7). We take this as further confirmation that the zero-field satellite $R_{xx}$ peaks in our measurement reflect changes in graphene band structure and formation of mini Dirac points, and are not Fabry-Perot resonance. $R_{yy}$ versus magnetic field (Fig. 4b) shows a similar overall trend as $R_{xx}$, however we note that $R_{xx}$ is about 2 orders of magnitude larger than $R_{yy}$, as the carrier wavefunction under magnetic field is localized along x direction, leading to a high $R_{xx}$. (Supplementary Information S6) . Both $R_{xx}$ and $R_{yy}$ show an overall resistance modulation, evolving along curved trajectories that follow a $B \propto \sqrt{n}$ relation. (dashed white lines in Fig. 4a,b). We interpret these features as commensurability oscillations (Fig. 4d)[32, 33]. In the semi-classical picture, the overall $R_{xx}$ resistance is expected to be $1/B$-periodic, with a minima resistance appearing whenever the cyclotron diameter matches the periodicity of the superlattice, i.e. when $2r_c$ satisfies

$$2r_c = \left(\lambda - \frac{1}{4}\right)L \tag{2}$$

with $\lambda$ being an integer[24, 34] and $L$ the lattice period. The locations of minima in our $R_{xx}$ data show excellent agreement with (2), consistent with the results previously reported by Drienovsky et al[24]. Alternatively, commensurability oscillations can be understood quantum mechanically by a $1/B$-periodic oscillation in the Landau level bandwidth, which reaches zero when (2) is satisfied. We note that the commensurability oscillations along the two measured directions appear out of phase such that when $R_{xx}$ fades, $R_{yy}$ grows, and vice versa. (Supplementary Information S6). The LL degeneracies are lifted by the 1D SL potential with each LL acquiring a dispersion in $k_y$ only. Thus, when the LL width in the direction of $k_y$ is large, $R_{xx}$ is large, since $R_{xx} \propto v_y^2$ where $v_y = \frac{1}{\hbar}\frac{dE}{dk_y}$. Conversely, $R_{yy} \propto v_x^2$ and so is less sensitive to variation of the intraband scattering and instead varies directly with the DOS, causing $R_{yy}$ to scale inversely with the level width. Therefore, zero LL bandwidth (zero dispersion in $k_y$, corresponding to the condition in (2) ) leads to an $R_{xx}$ minimum, simultaneous with an $R_{yy}$ maximum, consistent with the observation in Fig. 4a,b,d and providing further confirmation that the 1D superlattice imparts an asymmetric distortion of the bandstructure.

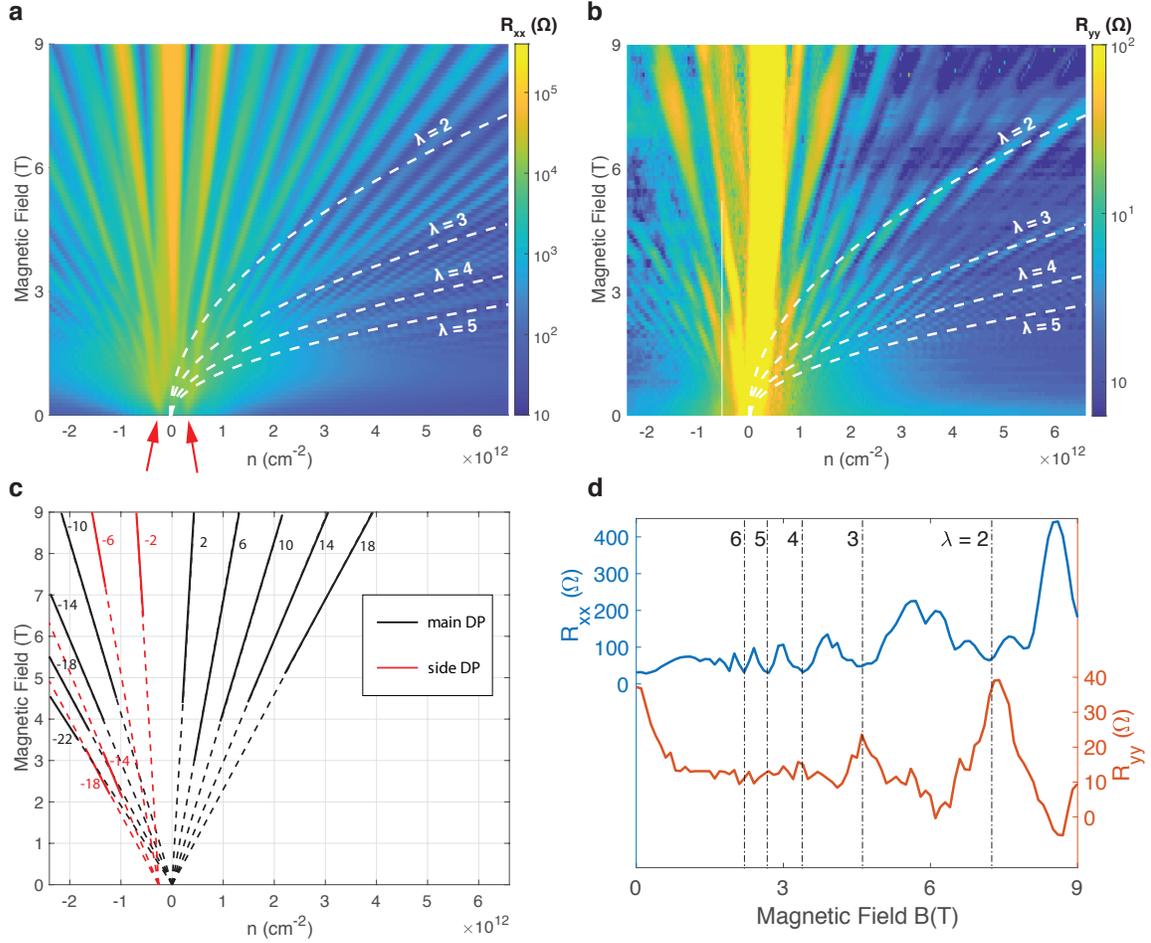

**Fig.4 | Magnetotransport in a 1D SL device. a,** Measured longitudinal resistance $R_{xx}$ as a function of carrier density $n$ and magnetic field $B$, in a $L = 47 nm$ device with $V_{SL} = 48V$. Red arrows indicate the density locations of the where satellite Landau fans converge. **b,** $R_{yy}$ measured from the same devices. In **a** and **b** dashed white lines trace along $2r_c = \left(\lambda - \frac{1}{4}\right)L$, with $\lambda$ being an integer and $r_c = \hbar\sqrt{\pi n}/eB$ the cyclotron radius (see text). **c,** Fan tracing diagram hightlighting traces of $R_{xx}$ minima from Fig. 4a. The $R_{xx}$ minima features associated with the main (satellite) Dirac point are colored in black (red), resepectively. Numbers indicate the filling fractions associated with the main or satellite fan. **d,** $R_{xx}$ and $R_{yy}$ as a function of magnetic field $B$ at carrier density $n = 6.53 \times 10^{12} cm^{-2}$. Dashed lines indicate the magnetic field at which the corresponding oscillation is theoretically expcted (see text). In order to suppress the quantum Hall oscilations and highlight the commensurability oscillations, the plotted curves are obtained by averaging the measured resistance over a small density window of $(6.53 \pm 0.23) \times 10^{12} cm^{-2}$.[see Supplementary Information S8]

In summary, we have presented a full characterization of transport anisotropy in a graphene 1D superlattice system. Our devices host a multitude of band structure features including mini Dirac points and anisotropic flattened Dirac points, the latter of which is not found in 2D systems engineered with 2D SL studied so far. The 1D SL gate acts as a powerful tuning knob in the system that cyclically flattens and unflattens a given Dirac point. Under finite magnetic field, the Landau level spectra are a mixture of main DP Landau levels, mini DP Landau levels, and Landau level dispersion in $k_y$ direction which gives rise to anisotropic commensurability oscillations. Our findings demonstrate a path to engineering transport asymmetry in a high mobility material, with full tunability of the anisotropy, and without the constraints of air sensitivity, such as has been encountered in the naturally corrugated phosphorene devices.[35] Patterning 2D systems beyond graphene using a periodic 1D SL[36, 37] could open new opportunities by coupling anisotropic bandstructures to properties not inherent in graphene such as strong spin-orbit coupling and magnetic ordering.

**Method:**

See Supplementary Information S1

**Acknowledgements:**

This research was supported primarily by the Office of Naval Research (ONR) Young Investors Program (no. N00014-17-1-2832). P.M. was supported by Science and Technology Commission of Shanghai Municipality grant no. 19ZR1436400, and NYU-ECNU Institute of Physics at NYU Shanghai. This research was carried out on the High Performance Computing resources at NYU Shanghai.

**References:**
1. Park, C.H., Yang, L., Son, Y.W., Cohen, M.L. & Louie, S.G. Anisotropic behaviours of massless Dirac fermions in graphene under periodic potentials. *Nat. Phys.* **4.,** 213-217 (2008).
2. Park, C.H., Son, Y.W., Yang, L., Cohen, M.L. & Louie, S.G. Landau levels and quantum Hall effect in graphene superlattices. *Phys. Rev. Lett.* **103,** 046808 (2009).
3. Brey, L. & Fertig, H.A. Emerging Zero Modes for Graphene in a Periodic Potential. *Phys. Rev. Lett.* **103,** 046809 (2009).

4.  Cao, Y. et al. Unconventional superconductivity in magic-angle graphene superlattices. *Nature* **556,** 43-50 (2018).
5.  Cao, Y. et al. Correlated insulator behaviour at half-filling in magic-angle graphene superlattices. *Nature* **556,** 80-84 (2018).
6.  Forsythe, C. et al. Band structure engineering of 2D materials using patterned dielectric superlattices. *Nat. Nanotechnol.* **13,** 566-571 (2018).
7.  Xia, F., Wang, H., Hwang, J.C.M., Neto, A.H.C. & Yang, L. Black phosphorus and its isoelectronic materials. *Nat. Rev. Phys.* **1,** 306-317 (2019).
8.  Shi, L.K., Ma, J. & Song, J.C.W. Gate-tunable flat bands in van der Waals patterned dielectric superlattices. *2D. Mater.* **7,** 015028 (2019).
9.  Geim, A.K. & Novoselov, K.S. The rise of graphene. *Nat. Mater.* **6,** 183-191 (2007).
10. Novoselov, K.S., Mishchenko, A., Carvalho, A. & Neto, A.H.C. 2D materials and van der Waals heterostructures. *Science* **353,** aac9439 (2016).
11. Dean, C.R. et al. Hofstadter's butterfly and the fractal quantum Hall effect in moire superlattices. *Nature* **497,** 598-602 (2013).
12. Ponomarenko, L.A. et al. Cloning of Dirac fermions in graphene superlattices. *Nature* **497,** 594-597 (2013).
13. Yankowitz, M. et al. Tuning superconductivity in twisted bilayer graphene. *Science* **363,** 1059-1064 (2019).
14. Sharpe, A.L. et al. Emergent ferromagnetism near three-quarters filling in twisted bilayer graphene. *Science* **365,** 605-608 (2019).
15. Yankowitz, M. et al. Emergence of superlattice Dirac points in graphene on hexagonal boron nitride. *Nat. Phys.* **8,** 382-386 (2012).
16. Hunt, B. et al. Massive Dirac Fermions and Hofstadter Butterfly in a van der Waals Heterostructure. *Science* **340,** 1427-1430 (2013).
17. Serlin, M. et al. Intrinsic quantized anomalous Hall effect in a moire heterostructure. *Science* **367,** 900-903 (2020).
18. Wang, P.L. & Xiong, S.J. Intrasubband Plasmons and Optical-Transmission in Random-Layer-Thickness N-I-P-I Semiconductor Superlattices. *Phys. Rev. B* **49,** 10373-10380 (1994).
19. Barbier, M., Vasilopoulos, P. & Peeters, F.M. Extra Dirac points in the energy spectrum for superlattices on single-layer graphene. *Phys. Rev. B* **81,** 075438 (2010).
20. Zhao, P.L. & Chen, X. Electronic band gap and transport in Fibonacci quasi-periodic graphene superlattice. *Appl. Phys. Lett.* **99,** 182108 (2011).
21. Dubey, S. et al. Tunable Superlattice in Graphene To Control the Number of Dirac Points. *Nano. Lett.* **13,** 3990-3995 (2013).
22. Drienovsky, M. et al. Towards superlattices: Lateral bipolar multibarriers in graphene. *Phys. Rev. B* **89,** 115421 (2014).
23. Drienovsky, M. et al. Few-layer graphene patterned bottom gates for van der Waals heterostructures. Preprint at http://arxiv.org/abs/1703.05631 (2017).
24. Drienovsky, M. et al. Commensurability Oscillations in One-Dimensional Graphene Superlattices. *Phys. Rev. Lett.* **121,** 026806 (2018).
25. Kuiri, M., Gupta, G.K., Ronen, Y., Das, T. & Das, A. Large Landau-level splitting in a tunable one-dimensional graphene superlattice probed by magnetocapacitance measurements. *Phys. Rev. B* **98,** 035418 (2018).
26. Dean, C. et al. Graphene based heterostructures. *Solid State Commun.* **152,** 1275-1282 (2012).
27. Wang, L. et al. One-Dimensional Electrical Contact to a Two-Dimensional Material. *Science* **342,** 614-617 (2013).

28. Dean, C.R. et al. Boron nitride substrates for high-quality graphene electronics. *Nat. Nanotechnol.* **5,** 722-726 (2010).
29. Allen, P.B. in Quantum Theory of Real Materials. (eds. J.R. Chelikowsky & S.G. Louie) viii, 549 p. (Kluwer Academic Publishers, Boston; 1996).
30. Madsen, G.K.H. & Singh, D.J. BoltzTraP. A code for calculating band-structure dependent quantities. *Comput. Phys. Commun.* **175,** 67-71 (2006).
31. Shore, K.A. Introduction to Graphene-Based Nanomaterials: From Electronic Structure to Quantum Transport. *Contemp. Phys.* **55,** 344-345 (2014).
32. Weiss, D., Vonklitzing, K., Ploog, K. & Weimann, G. Magnetoresistance Oscillations in a Two-Dimensional Electron-Gas Induced by a Submicrometer Periodic Potential. *Europhys. Lett.* **8,** 179-184 (1989).
33. Gerhardts, R.R., Weiss, D. & Vonklitzing, K. Novel Magnetoresistance Oscillations in a Periodically Modulated Two-Dimensional Electron-Gas. *Phys. Rev. Lett.* **62,** 1173-1176 (1989).
34. Beenakker, C.W.J. Guiding-Center-Drift Resonance in a Periodically Modulated Two-Dimensional Electron-Gas. *Phys. Rev. Lett.* **62,** 2020-2023 (1989).
35. Qiao, J.S., Kong, X.H., Hu, Z.X., Yang, F. & Ji, W. High-mobility transport anisotropy and linear dichroism in few-layer black phosphorus. *Nat. Commun.* **5** (2014).
36. Wu, S., Killi, M. & Paramekanti, A. Graphene under spatially varying external potentials: Landau levels, magnetotransport, and topological modes. *Phys. Rev. B* **85,** 195404 (2012).
37. Xu, H. et al. Oscillating edge states in one-dimensional MoS2 nanowires. *Nat. Commun.* **7** (2016).